\pdfoutput=1%
\documentclass[a4paper,american,aip,citeautoscript,floatfix,jcp,pdftex,reprint,superscriptaddress,twocolumn]{revtex4-1}

\usepackage{amsfonts,amsmath,amssymb}%
\usepackage{graphicx}%
\usepackage[utf8]{inputenc}%
\usepackage[version=3]{mhchem}
\usepackage[per-mode=symbol]{siunitx}
\usepackage{textcomp}%
\usepackage[textsize=tiny]{todonotes}%
\usepackage{xspace}
\usepackage[colorlinks,citecolor=black,urlcolor=blue]{hyperref}%
\usepackage{hypernat}%

\newlength{\figwidth}%
\newlength{\figwidthlarge}%
\newlength{\figwidthsmall}%
\newcommand{\cost}{\ensuremath{\left\langle\cos^2\theta_\text{2D}\right\rangle}\xspace}

\newcommand{\degree}{\ensuremath{^\circ}\xspace}%
\newcommand{\eg}{e.\,g.}%
\newcommand{\Estat}{\ensuremath{\textup{E}_\textup{s}}\xspace}
\newcommand{\Estatabs}{\ensuremath{\text{E}_{\textup{s}}}}%
\newcommand{\expected}[1]{\left\langle #1\right\rangle}
\newcommand{\ie}{i.\,e.}%
\newcommand{\Iyag}{\ensuremath{\textup{I}_\textup{YAG}}\xspace}
\newcommand{\Iprobe}{\ensuremath{\textup{I}_\textup{probe}}\xspace}

\newcommand{\mueff}{\ensuremath{\mu_{\text{eff}}}}
\newcommand{\Nuptot}{\ensuremath{\text{N}_{\textup{up}}/\text{N}_{\textup{total}}}}

\DeclareSIUnit\intensity{\watt\per\centi\meter\squared}
\DeclareSIUnit\fieldstrength{\volt\per\centi\meter}
\DeclareSIUnit\kfieldstrength{k\volt\per\centi\meter}

\let\orgautoref\autoref

\renewcommand{\autoref}{%
  \def\equationautorefname{Eq.}%
  \def\figureautorefname{Fig.}%
  \def\subfigureautorefname{Fig.}%
  \def\tableautorefname{Tab.}%
  \orgautoref}

\newcommand{\inano}{\affiliation{Interdisciplinary Nanoscience Center (iNANO), Aarhus University,
      8000 Aarhus C, Denmark}}%
\newcommand{\granada}{\affiliation{Instituto Carlos I de F\'{\i}sica Te\'orica y Computacional and
      Departamento de F\'{\i}sica At\'omica, Molecular y Nuclear, Universidad de Granada, 18071
      Granada, Spain}}%
\newcommand{\aarchem}{\affiliation{Department of Chemistry, Aarhus University, 8000 Aarhus C,
      Denmark}}%
\newcommand{\cfel}{\affiliation{Center for Free-Electron Laser Science, DESY, 22607 Hamburg,
      Germany}}%
\newcommand{\cui}{\affiliation{The Hamburg Center for Ultrafast Imaging, University of Hamburg, 22761 Hamburg, Germany}}%
\newcommand{\uhh}{\affiliation{Department of Physics, University of Hamburg, 22761 Hamburg,
      Germany}}%

\begin{document}
\title{Mixed-field orientation of molecules without rotational symmetry}
\author{Jonas L.\ Hansen}\inano%
\author{Juan J.\ Omiste}\granada%
\author{Jens H. Nielsen}\aarchem%
\author{Dominik Pentlehner}\aarchem%
\author{Jochen K\"upper}\cfel\uhh\cui%
\author{Rosario Gonz\'{a}lez-F\'{e}rez}\granada\cui
\author{Henrik Stapelfeldt}\inano\aarchem%
\date{\today}%
\begin{abstract}\noindent%
   The mixed-field orientation of an asymmetric-rotor molecule with its permanent dipole moment non-parallel to the principal axes of polarizability is investigated experimentally and theoretically.
   We find that for the typical case of a strong, nonresonant laser field and a weak static electric field complete 3D orientation is induced if the laser field is elliptically polarized and if its major and minor polarization axes are not parallel to the static field. For a linearly polarized laser field solely the dipole moment component along the most polarizable axis of the molecule is relevant resulting in 1D orientation even when the laser polarization and the static field are non parallel.
   Simulations show that the dipole moment component perpendicular to the most-polarizable axis becomes relevant in a strong dc electric
   field combined with the laser field. This offers an alternative approach to 3D orientation by combining a linearly-polarized laser field and a strong dc electric field arranged at an angle equal to the angle between the most polarizable axis of the molecule and its permanent dipole moment.
\end{abstract}
\maketitle%

\section{Introduction}
The ability to control the rotational motion and to angularly confine molecules has various
applications in molecular sciences. This includes studies of steric effects in chemical reactions,
both bimolecular and photoinduced, and the possibility to investigate molecules from their own
point of view, the molecular frame. The latter mitigates the usual blurring of experimental
observables caused by the random orientation of molecules in uncontrolled samples. Access to
molecular frame measurements is crucial in several applications, notably in various modern schemes
aiming at observing the (coupled) motion of nuclei and electrons during chemical reactions~\cite{bisgaard_time-resolved_2009, Filsinger:PCCP13:2076, Sciaini:2011hi, hansen_control_2012,
   hensley_imaging_2012, Ullrich:ARPC63:635, Barty:2013ib}.

Methods based on the use of moderately intense, nonresonant, near-infrared laser pulses have proven
particularly useful for controlling the alignment and, in conjunction with weak dc electric fields,
orientation of a broad range of molecules. Alignment refers to
the confinement of molecule-fixed axes along laboratory-fixed axes, and orientation usually refers to the
molecular dipole moment pointing in a particular direction~\cite{stapelfeldt_colloquium:_2003}. For a linear molecule, only
a single axis needs to be confined in space to ensure complete rotational control. This can be
achieved by a linearly polarized laser pulse, which will align the most polarizable axis (MPA),
i.e. the internuclear axis of the molecule. This is termed one-dimensional (1D)
alignment \cite{sym-top}. Combined with a (weak) static electric field it can also control the head-versus-tail
order of a polar molecule, \ie, induce 1D orientation~\cite{friedrich_enhanced_1999,
   sakai_controlling_2003, buck_oriented_2006, holmegaard_laser-induced_2009,
   ghafur_impulsive_2009, rouzee_optimization_2009}.

Complete rotational control of asymmetric top molecules requires the confinement of three molecular
axes to laboratory frame fixed axes, resulting in 3D alignment. In the adiabatic limit, where the
laser pulse is turned on slower than the rotational periods of the molecule, it has been shown that
an elliptically polarized laser pulse can induce 3D alignment~\cite{larsen_three_2000,
   tanji_three-dimensional_2005, nevo_laser-induced_2009}. For polar molecules, where the permanent
dipole moment (DM) is parallel to the MPA it has also been shown that 3D orientation, defined as 3D
alignment and a unique direction of the DM, can be achieved by combining the elliptically polarized
laser pulse with a weak static electric field parallel to the major polarization axis~\cite{tanji_three-dimensional_2005, nevo_laser-induced_2009}. For most asymmetric top molecules, the
DM does, however, not coincide with any principal axis of polarizability. While 3D alignment is expected to work well for these
less symmetric molecules, it remains to be explored if the combined action of a linearly or
elliptically polarized laser pulse and a weak or strong static electric field can efficiently induce
3D orientation. Also, as discussed here, the precise meaning of 3D orientation must be specified for molecules with low symmetry.

In the current work we investigate 3D alignment and orientation of asymmetric top molecules where
the DM is not parallel to a principal axis of polarizability. Our studies are motivated by the fact that many important
biomolecules, \eg, amino acids, nucleic acids, peptides, and DNA strands, belong to this class of
molecules. Controlling how they are turned in space would be of significant value in novel and
emerging schemes for time-resolved molecular imaging~\cite{spence_single_2004, Barty:2013ib, holmegaard_photoelectron_2010}. Following the conclusions from the current work, this
three-dimensional control is indeed possible. Our studies focus on 6-chloropyridazine-3-carbonitrile
(\ce{C4N2H2ClCN}, CPC). The molecule is chosen because the DM is off-set by 57.1$\degree$ from the
MPA and because the atomic composition makes it possible to determine its 3-dimensional spatial
orientation through Coulomb explosion imaging.

\section{Molecular structure and electrical properties of CPC}
\begin{figure}
   \centering
   \includegraphics[width=\linewidth]{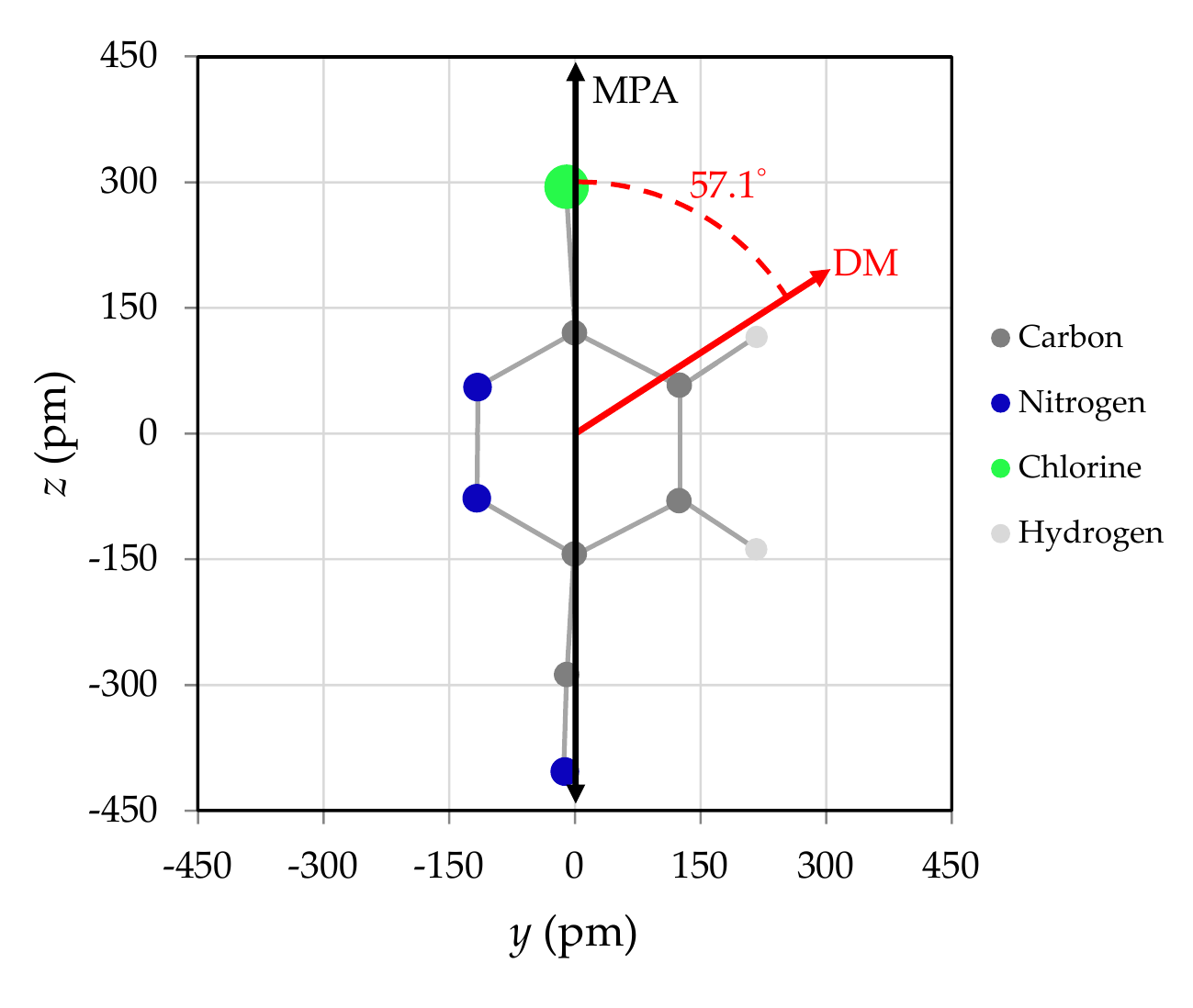}
   \caption{Sketch of the molecular structure of 6-chloropyridazine-3-carbonitrile with the most
      polarizable axis (black doubleheaded arrow) and permanent dipole moment (red arrow). The
      coordinates of the individual atomic positions in CPC, from a geometry optimized quantum
      chemical calculation, show a slight shift of the chlorine and nitrile bond angles towards the
      nitrogens in the pyridazine ring. The x-axis is perpendicular to the figure plane.}
   \label{fig:Mol-Structure}

 \end{figure}

A sketch of the molecular structure of CPC and the position of the MPA and DM vector are shown in
\autoref{fig:Mol-Structure}.
 The planar molecule consists of an aromatic pyridazine ring with a nitrile and a chlorine
 substituent. The N-N bond is shorter than the C-N and C-C bonds in the aromatic ring causing the
 bond angle of the substituents on the ring to bend slightly towards the pyridazine nitrogens, as
 can be seen from the energy optimized geometry of the molecule shown in
 \autoref{fig:Mol-Structure}. Quantum chemical calculations~\footnote{Gaussian
    2003~\cite{Gaussian:2003C02} B3LYP/aug-pc-1 calculations performed by Frank Jensen, Department
    of Chemistry, Aarhus University.} determine the electric dipole moment of CPC to be \SI{5.21}{D}
 with $\mu_{x}=\SI{0}{D}$, $\mu_{y}=\SI{4.37}{D}$, $\mu_{z}=\SI{2.83}{D}$, \ie, in the molecular
 plane and at an angle of 57.1\,\degree with respect to the $z$ axis. The static polarizability
 components of CPC are determined to be $\alpha_{xx}=\SI{7.88}{\AA^3}$,
 $\alpha_{yy}=\SI{12.0}{\AA^3}$ and $\alpha_{zz}=\SI{22.3}{\AA^3}$.

\section{Experimental Setup}
Most aspects of the experimental setup have been described previously~\cite{filsinger_quantum-state_2009, nielsen_stark-selected_2011, hansen_control_2012} and only a few
pertinent details will be given here. A few mbar of CPC (ChemFuture PharmaTech, $> \SI{97}{\%}$
chemical purity) was seeded in a helium carrier gas at a backing pressure of \SI{90}{bar} and
expanded into vacuum through a pulsed Even-Lavie valve~\cite{even_cooling_2000} heated to
\SI{170}{\degree} C. The expansion was skimmed twice before entering an electrostatic deflector, where
the molecules were deflected according to the effective dipole moment \mueff\ of their specific
rotational quantum state~\cite{filsinger_quantum-state_2009}.

The quantum-state-dispersed molecular beam entered a velocity map imaging (VMI) spectrometer where
it was crossed at \SI{90}{\degree} by two collinear laser beams. The laser beams were focused by a
spherical lens ($f=\SI{30}{cm}$) mounted on a motorized translation stage. This allowed for the
height of the foci to be scanned with high precision. Hereby it is possible to measure the vertical intensity profile of the molecular beam such that the effect of the electrostatic deflector can be characterized. Furthermore, the focused laser beams can be directed to the most deflected molecules which are the ones residing in the lowest lying rotational states and, thereby, those that undergo the strongest alignment and orientation~\cite{filsinger_quantum-state_2009}. The molecules were aligned and oriented by the
combined effect of pulses from one of the laser beams (YAG pulse, $\lambda=\SI{1064}{nm}$,
$\tau_\text{FWHM}=\SI{10}{ns}$, $\omega_0=\SI{34}{\um}$, $\Iyag=\SI{8e11}{\intensity}$, injection
seeded) and the weak static electric field from the VMI spectrometer (\Estat was varied between
values of \SI{571}{V/cm} and \SI{714}{V/cm}). The polarization of the YAG pulse can be rotated
to rotate the molecular alignment with respect to the static field direction which is fixed by the VMI spectrometer axis -- see Sec.~\ref{sec:orientation}.
The YAG beam was overlapped in space and
time with pulses from a second laser (probe pulse: \SI{800}{nm}, \SI{30}{fs}, \SI{24}{\um},
$\Iprobe=\SI{4e14}{\intensity}$). These short pulses multiply ionized the molecules, which then
fragmented into charged ions. The ions were projected onto a 2-dimensional particle detector in order
to detect their recoil directions. For CPC molecules, \ce{Cl+} ion momenta were recorded to
determine the spatial orientation of the \ce{C-Cl} bond axis with respect to the laboratory frame.
The \ce{N+} or \ce{H+} fragment ion distributions were recorded to provide information about the
orientation of the molecular plane in the laboratory frame. All experiments were conducted on
deflected, state-selected molecular samples at a repetition rate of \SI{20}{Hz}, limited by the YAG
laser.

\section{Theorerical description}
Theoretically we investigated the rotational dynamics of the CPC molecule in combined static
electric and nonresonant laser fields polarized either linearly or elliptically. Due to the complexity of
this system, we retreated to a
quasi-static description. We assumed that the interaction with the laser pulse can be described
within the adiabatic limit. We applied a two-photon rotating-wave approach averaging over the rapid
oscillations of the nonresonant field. In the framework of the rigid-rotor approximation, we solved
the time-independent Schrödinger equation of the CPC molecule in a field configuration equivalent to
the experimental one. It is convenient to define a laser-polarization-fixed frame (LPFF) $(X,Y,Z)$. For the elliptically
polarized field, the major polarization axis defines the $Z$-axis and the minor polarization the $Y$-axis.
For the linearly polarized field, the $Z$-axis is defined by the polarization axis. The homogeneous electrostatic field of strength
$\text{E}_{\textup{s}}$ is contained in the $YZ$-plane at an angle $\beta$ with respect to the $Z$
axis. The relation between the LPFF and the molecular fixed frame (MFF) $(x,y,z)$ (defined in \autoref{fig:Mol-Structure}) is given by the Euler
angles $\Omega=(\phi,\theta,\chi)$~\cite{zare_angular_1988}. The Hamiltonian of this system is
\begin{equation}
   H = J_{x}^2B_{x}+J_{y}^2B_{y}+ J_{z}^2 B_{z} +H_s + H_l
   \label{eqn:hamiltonian}
\end{equation}
with the rotational constant $B_{x}$, $B_{y}$ and $B_{z}$ and the interaction operators $H_s$ and
$H_l$ with the dc and ac electric fields, respectively.

The Stark interaction reads
\begin{align}
   \label{eqn:hamil_electric} H_{s} = &-\mathbf{E}_s\cdot\boldsymbol{\mu} \\
   & =- \text{E}_{\textup{s}}\mu\cos\theta_{s\mu} \nonumber\\
   & = - \text{E}_{\textup{s}}\mu_z\cos\theta_{sz}-\text{E}_{\textup{s}}\mu_y\cos\theta_{sy} \nonumber
\end{align}
with the absolute value of the electric dipole moment $\mu$, and its two
components $\mu_z$ and $\mu_y$. The angles between the electric field and $\boldsymbol{\mu}$, and the MFF $z$ and
$y$-axes, $\theta_{s\mu}$, $\theta_{sz}$ and $\theta_{sy}$, respectively, are given by the relations
\begin{align}
   \label{eqn:cos_sz} \cos\theta_{sz} =& \cos\beta\cos\theta+\sin\beta\sin\theta\sin\phi, \\
      \cos\theta_{sy} =& \cos\beta\sin\theta\sin\chi \nonumber\\
   &    \label{eqn:cos_sy} +\sin\beta(\cos\phi\cos\chi-\cos\theta\sin\phi\sin\chi) \\
   \label{eqn:cos_smu} \cos\theta_{s\mu} =& \cos(57.1\degree)\cos\theta_{sz}+\sin(57.1\degree)\cos\theta_{sy}.
\end{align}

The interaction of the molecule with a nonresonant elliptically polarized laser field can be written
as
\begin{align}
   H_{l} =
   &-\cfrac{\text{I}_{ZZ}}{2c\epsilon_0}\left(\alpha^{zx}\cos^2\theta_{Zz}+\alpha^{yx}\cos^2\theta_{Zy}\right) \nonumber\\
   \label{eqn:hamil_laser_ellip} & -\cfrac{\text{I}_{YY}}{2c\epsilon_0}\left(\alpha^{yx}\cos^2\theta_{Yy}+\alpha^{zx}\cos^2\theta_{Yz}\right)
\end{align}
where $\text{I}_{YY}$ and $\text{I}_{{ZZ}}$ are the intensities of the polarization components along
the LPFF $Y$ and $Z$ axes, respectively. The total intensity is $\Iyag=\text{I}_{YY}+\text{I}_{ZZ}$,
and $\text{I}_{ZZ}=3\text{I}_{YY}$ is used here. $\alpha^{ji}=\alpha_{jj}-\alpha_{ii}$, and
$\alpha_{ii}$ are the $i$-th diagonal element of the polarizability tensor, with $i=x,y,z$.
$\epsilon_0$ the dielectric constant and $c$ is the speed of light. $\theta_{Pq}$ are the angles
between the LPFF $P$-axis and the MFF $q$-axis, and they are related to the Euler angles as follows
\begin{align*}
  \cos\theta_{Zz} &= \cos\theta,\\
  \cos\theta_{Zy} &= \sin\theta\sin\chi,\\
  \cos\theta_{Yz} &= \sin\phi\sin\theta,\\
  \cos\theta_{Yy} &= \cos\phi\cos\chi-\cos\theta\sin\phi\sin\chi.
\end{align*}
If the laser field is linearly polarized, the interaction with this field is obtained by setting
$\text{I}_{YY}=0$ and  $\Iyag=\text{I}_{ZZ}$ in \autoref{eqn:hamil_laser_ellip}.

The time-independent Schrödinger equation of the Hamiltonian \autoref{eqn:hamiltonian} was solved by
expanding the wave function in a basis set formed by linear combinations of field-free symmetric top
wave functions, i.e. Wigner functions~\cite{zare_angular_1988}. For each field
configuration, we constructed a basis that respects the symmetries of the corresponding irreducible
representation~\cite{omiste:064310}.

Let us shortly summarize the symmetries of this system in the mixed-field configurations. In the
field-free case, they are given by the spatial group SO(3) and the molecular point group
D$_2$~\cite{kanya_pendular-limit_2004, omiste:064310}.
As a consequence, the total angular momentum $J$ and its
projection $M$ onto the $Z$-axis of the LPFF are good quantum numbers, but the projection of $J$ onto the
$z$-axis of the MFF ($K$) is not well defined.
The symmetries of the  Hamiltonian \autoref{eqn:hamiltonian} with a linearly polarized laser
and a dc electric field tilted by an angle $\beta$ have been analyzed in detail in
Ref.~\onlinecite{omiste:064310}.
For an elliptically polarized laser field in
the $YZ$ plane and with the dc field parallel to the $Z$-axis, \ie, $\beta=180\degree n$, a $\pi$-rotation
around the LPFF $Z$-axis and the reflection on the $YZ$-plane (the laser polarization plane)
are the
symmetry operations and $M$ is not a good quantum number, but its parity is. For the other two
cases, $\beta=90\degree (2n+1)$ and $\beta\ne{}180\degree n$, the system has the same
symmetries as in the
corresponding field configuration with a linearly polarized laser field,
see Ref.~\onlinecite{omiste:064310}.

\section{Experimental  Results}
\subsection{Alignment}
\begin{figure}
  \centering
  \includegraphics{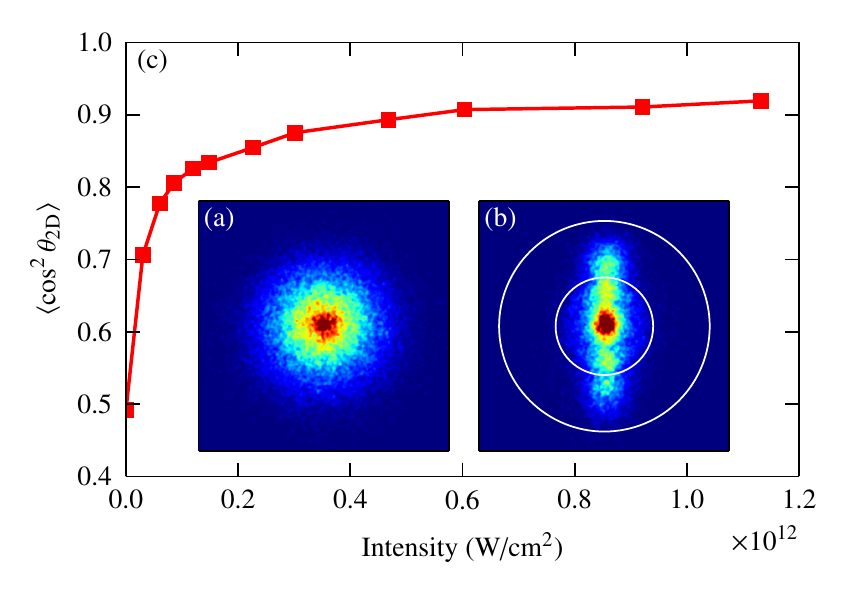}
  \caption{\ce{Cl+} images recorded (a) without and (b) with the YAG pulse,
     $\Iyag=\SI{1.1e12}{\intensity}$. (c) The degree of alignment \cost as a function of \Iyag.}
  \label{fig:1D-alignment}
\end{figure}

We start by showing that a linearly polarized YAG pulse induces 1D alignment of the CPC molecules. For
this purpose, the emission directions of \ce{Cl+} ions are detected. The expected action of the YAG pulse
is that it aligns the MPA along its polarization axis and as such an experimental observable that
provides direct and precise information about the spatial orientation of this axis would be ideal.
Unlike in higher-symmetry molecules, \eg, iodobenzene~\cite{holmegaard_laser-induced_2009}, no such
observable exists. The emission direction of \ce{Cl+} ions comes close, assuming axial recoil along
the C-Cl bond axis, since the C-Cl axis is only off-set by 3 degrees from the MPA. When only the
linearly-polarized probe pulse is applied, polarized perpendicular to the detector, the \ce{Cl+}
image shown in \autoref{fig:1D-alignment}\,(a) is circularly symmetric, as expected for randomly
oriented molecules.
When the YAG pulse is included the \ce{Cl+} ions tightly localize along its polarization axis parallel to
the detector plane, see \autoref{fig:1D-alignment}\,(b). These observations show that the C-Cl bond
axes of the CPC molecules are aligned along the YAG pulse polarization axis, \ie, that 1D alignment is
induced. The degree of alignment is quantified by determining the average value of
$\cos^2\theta_\text{2D}$, \cost, where $\theta_\text{2D}$ is the angle between the YAG pulse
polarization and the projection of a \ce{Cl^+} ion velocity vector on the detector screen. Only a
confined radial range is used to determine \cost. This range at the outermost part of the images is
marked by circles in \autoref{fig:1D-alignment}\,(b). It corresponds to ions originating from a
highly directional Coulomb explosion process. The derived values are plotted as a function of
the YAG pulse intensity, \Iyag in \autoref{fig:1D-alignment}(c). \cost rises from 0.5, the value
characterizing a sample of randomly oriented molecules, at $\Iyag=\SI{0}{\intensity}$
to $0.93$ at the highest value of
\Iyag. This behaviour is fully consistent with many previous studies of 1D adiabatic
alignment~\cite{stapelfeldt_colloquium:_2003, holmegaard_laser-induced_2009}. The \cost
values determined underestimate the true degree of alignment due to the offset between the C-Cl axis and the MPA and imperfect axial recoil.

\begin{figure}[t]
  \centering
  \includegraphics{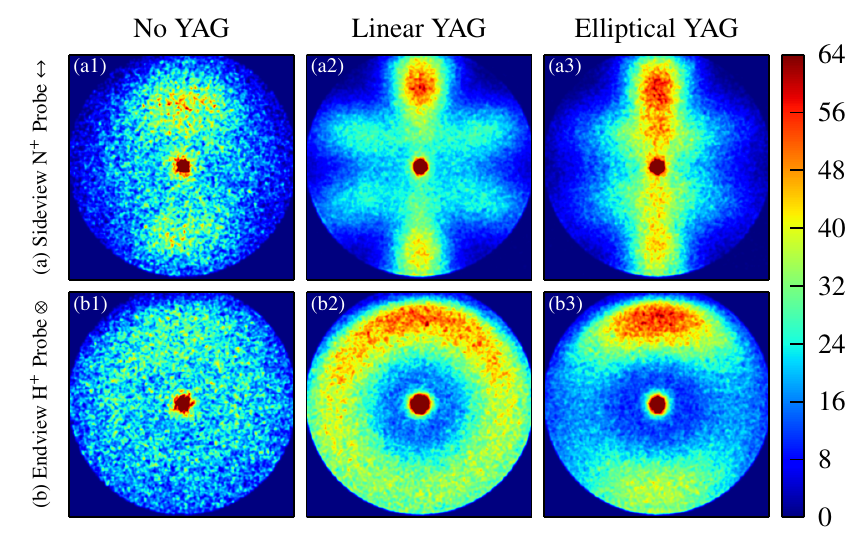}.
  \caption{a) \ce{N+} and b) \ce{H+} images demonstrating 1D and 3D alignment of CPC. The color
     scale is chosen such that the central low kinetic energy peak is saturated to enhance the
     visibility of the rest of the image. The mild up-own asymmetry observed in the images is caused by a slightly reduced detection
     efficiency on the lower part of the detector.}
  \label{fig:cpc-alignment}
\end{figure}

In order to investigate the effect on the molecular alignment when the YAG pulse polarization is changed
from linear to elliptical, an ellipticity ratio of 3:1 was applied, \ie, the intensity along
the major polarization axis of the YAG pulse is three times the intensity along the minor axis. For these
measurements, \ce{N^+} and \ce{H^+} images, displayed in \autoref{fig:cpc-alignment}, are used to infer
information about the molecular alignment.
The images represent either a "side-view" when the major polarization axis is parallel to the
detector (vertical in \autoref{fig:cpc-alignment}), \ie, the molecules are watched from the side, or an "end-view" when the major polarization
axis is perpendicular to the detector, \ie, the molecules are watched from the end~\cite{viftrup_holding_2007}.

For \ce{N+} ions, the molecule is imaged in side-view. \autoref{fig:cpc-alignment}\,(a1) shows the
image obtained with the probe pulse by itself, polarized vertically, and serves as a reference. In panel (a2) measurements
including the linearly polarized YAG pulse are shown. The \ce{N+} ions appear as two distinct areas at
large radii along the vertical axis and as two pairs of wings protruding nearly horizontally from
the vertical centerline. The 1D alignment means that the MPA of the molecules is confined along the vertical $Z$ axis
and it implies that the recoiling \ce{N+} ions from the \ce{CN} group will be ejected vertically, either up
or down depending on the orientation of the molecule. These ions form the two distinct centerline regions
of signal similar to the \ce{Cl+} ion structure (\autoref{fig:1D-alignment}(b))
used for the determination of the 1D alignment discussed above. The wing structure is interpreted as
\ce{N+} ions originating from Coulomb explosion of the N atoms in the aromatic ring. Since the
linearly polarized YAG pulse does not impose any constraint on the rotation of the ring the \ce{N+} ions
will be emitted in a double-torus-like pattern. Upon projection on the 2D detector plane this gives
the wing-structure. Covariance analysis~\cite{hansen_control_2012} confirms that in the
wing-structure the two \ce{N+} ions from a single molecule are predominately produced on the same
sides of the tori supporting this interpretation. When the YAG pulse polarization is changed to elliptical
the image in \autoref{fig:cpc-alignment}~(a3) is obtained. The \ce{N+} ions from the ring are confined
close to the vertical axis, whereas the \ce{N+} ion emission structure from the CN group is
practically unchanged. This shows that the alignment of the MPA is not much changed while the molecular
plane is no longer free to rotate, but instead it is confined to the polarization plane. This
demonstrates that the molecule is 3D aligned. The corresponding end-view images of \ce{H+} in row~(b)
corroborate this interpretation: With a linearly polarized YAG pulse the \ce{H+} ions emerge in the
circularly symmetric pattern shown in panel (b2), corresponding to free rotation of the molecular
plane around the YAG pulse polarization axis. (Note that in this measurement the probe pulse polarization is perpendicular to the detector plane). For an elliptically polarized YAG pulse, the \ce{H+} ions
are angularly localized around the vertical minor polarization axis, \ie, the molecular plane is
confined to the polarization plane. The radial structures of panel (b2) and (b3) are the same,
confirming that the long axes of the molecules remain aligned along the major polarization axis.
Thus, the \ce{H+} images confirm that the CPC molecules are 3D aligned by the elliptically polarized
YAG pulse.

\subsection{Orientation}
\label{sec:orientation}
\begin{figure}
  \centering
  \includegraphics[width=\columnwidth]{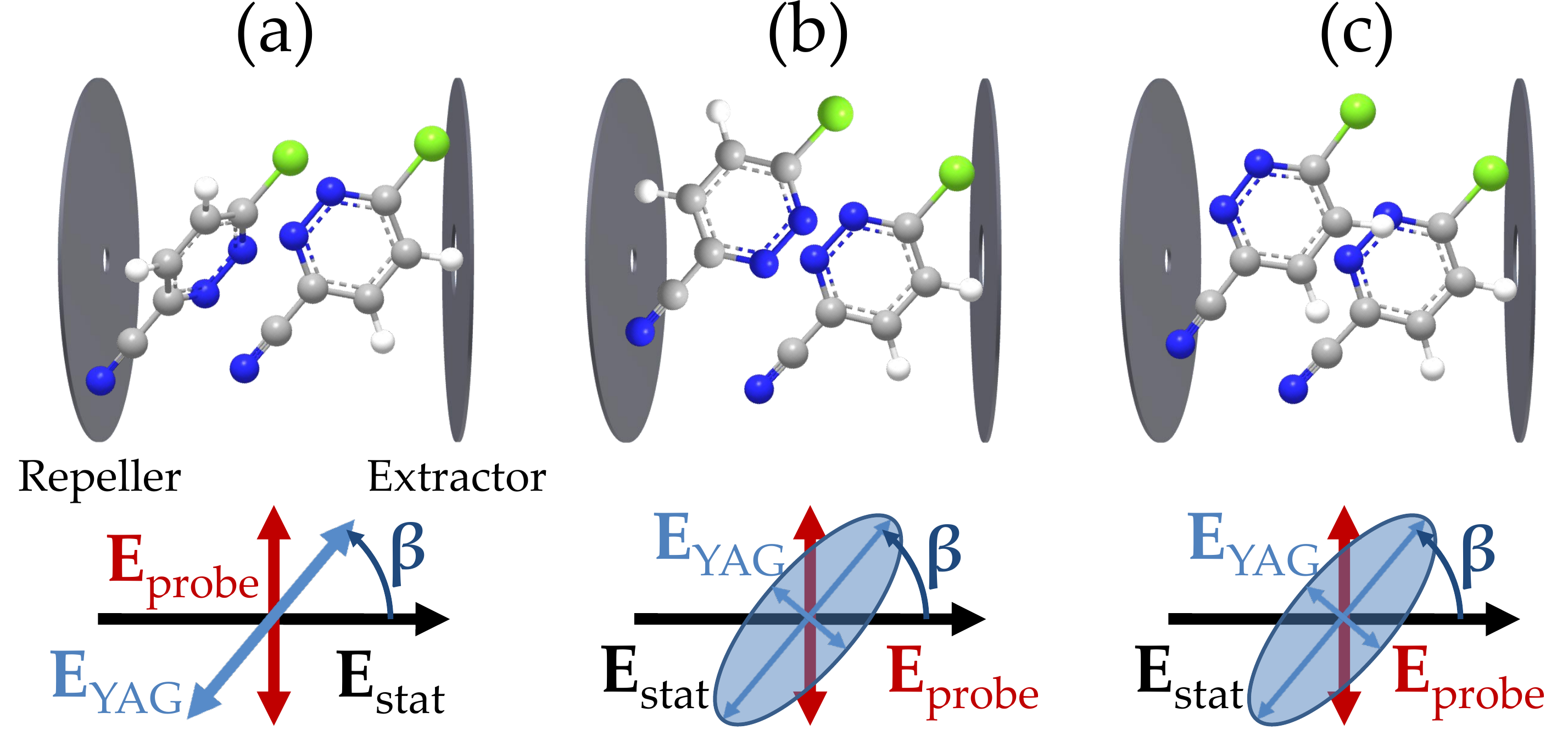}
  \caption{Schematic illustration of (a): 1D orientation; (b): 1D orientation of the MPA plus 3D alignment; (c): 3D orientation.
     The polarization state of the YAG pulse and the probe pulse with respect to the static electric field and the detector plane is shown below each of the 3 molecular sketches.
     In the orientation experiments the polarization direction of the probe pulse is kept fixed in the
     plane of the detector, while $\beta$ (the angle between the static field direction and the
     (major) polarization axis of the YAG pulse) is changed.}
  \label{fig:CPC_exp_setup_1D_3D_orient}
\end{figure}

First we discuss the orientation that results from the combined action of a linearly polarized, moderately intense laser and a weak static electric field. All previous studies comprised molecules where the permanent dipole moment was parallel to the MPA including linear rotors~\cite{sakai_controlling_2003, buck_oriented_2006, nielsen_laser-induced_2012}, symmetric tops~\cite{ghafur_impulsive_2009, rouzee_optimization_2009} and asymmetric tops~\cite{holmegaard_laser-induced_2009, holmegaard_photoelectron_2010}. In these cases 1D orientation, defined as 1D alignment and a preferred direction of the permanent dipole moment, was induced. For studies employing ion imaging, as in the current work, orientation was
observed when the molecule was rotated away from the side-view geometry used in pure alignment
measurements; see \autoref{fig:CPC_exp_setup_1D_3D_orient}(a) for a sketch of this experimental
approach.
In practice this was done by rotating the YAG pulse polarization to angles where $\beta \neq
\SI{90}{\degree}$. This provides a component of the static field along the dipole moment which mixes
the pendular states of the tunneling doublet to form the corresponding oriented states. The
experimental findings showed that the degree of orientation increased monotonically as $\beta$ was
rotated from $\SI{90}{\degree}$ towards $\SI{0}{\degree}$ or $\SI{180}{\degree}$~\cite{holmegaard_laser-induced_2009}. Later experiments
and analysis have identified this behaviour as resulting from nonadiabatic dynamics in the
mixed-field orientation \cite{nielsen_laser-induced_2012, nielsen_making_2012,
   omiste_nonadiabatic_2012,PhysRevA.88.033416}. In the following we investigate if the angular offset of the
dipole moment from the MPA in CPC influences the efficiency of mixed-field orientation and if the
degree of orientation peaks when the MPA or the permanent dipole moment is directed along the static
field from the VMI spectrometer. The experimental observables used are the \ce{Cl+} ion images which
provide information about the orientation of the MPA.

\begin{figure}[t]
  \centering
  \includegraphics{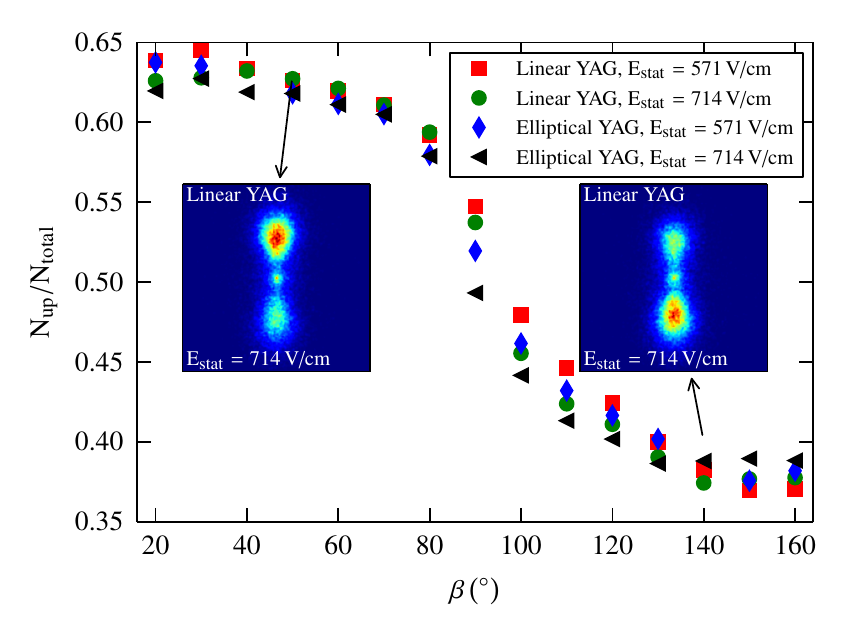}
  \caption{Degree of orientation as a function of the angle $\beta$ between the static field and the (major)
     polarization axis of the YAG pulse for different field strengths. The values are determined from the distribution of
     \ce{Cl+} ions. The insets show raw ion images at (left) $\beta=\SI{50}{\degree}$ and (right)
     $\beta=\SI{140}{\degree}$.}
  \label{fig:beta}
\end{figure}

Examples of \ce{Cl+} ion images recorded for $\beta=\SI{50}{\degree}$ $(\SI{140}{\degree})$ are
shown as insets in \autoref{fig:beta}.
For $\beta=\SI{50}{\degree}$~$(\SI{140}{\degree})$, more (less) ions are detected on the upper half
of the detector than on the lower half. In analogy with previous studies we interpret these
observations as 1D orientation due to the combined effect of the YAG laser field and the static
electric extraction field~\cite{holmegaard_laser-induced_2009}, where the DM $z$-component orients
along the projection of the dc electric field onto the MPA. This implies that the partially negative
nitrile end will be directed towards the repeller electrode where
the potential is highest and the C-Cl bond towards the extractor electrode where the potential is lowest
-- see \autoref{fig:CPC_exp_setup_1D_3D_orient}(a). As a consequence, \ce{Cl+} ions are expected to be
ejected upwards (downwards) for $\beta=\SI{50}{\degree}$~$(\SI{140}{\degree})$. This is in agreement
with the up-down asymmetry in the images.

The degree of 1D orientation is quantified by dividing the number of ions detected on the upper half of
the detector by the total amount of ions detected (\Nuptot), corrected for the slight up-down detector asymmetry. This asymmetry parameter, is plotted in
\autoref{fig:beta} as a function of $\beta$ recorded for two values of the static electric field. The figure shows
that the degree of orientation increases gradually as the MPA is rotated
towards the direction of the static electric field. This behaviour is similar to that observed for molecules such as OCS~\cite{holmegaard_photoelectron_2010} and iodobenzene~\cite{holmegaard_laser-induced_2009}
where the MPA and the permanent dipole moment are parallel. The experimental findings are rationalized by our
computational treatment -- discussed in \autoref{sec:theory-results}.

Next we investigated 3D orientation. In previous studies 3D orientation was defined as 3D alignment occurring together
with a preferred direction of the permanent dipole moment axis, which coincides with the MPA for molecules with rotational point group symmetry like iodobenzene and benzonitrile (molecular point group $C_{2v}$). In the case of molecules like CPC (molecular point group $C_s$) full 3D orientation [\autoref{fig:CPC_exp_setup_1D_3D_orient}(c)] requires preferred directions in space for at least two components of the dipole moment. For CPC this would be the $z$-axis and the $y$-axis -- see \autoref{fig:Mol-Structure}. A slightly less complete 3D confinement of the molecular rotations is 3D alignment and 1D orientation of one of the principal axes of polarizability. This can be either the $z$-axis (the MPA) as illustrated in \autoref{fig:CPC_exp_setup_1D_3D_orient}(b) or the $y$-axis which will be discussed in \autoref{sec:theory-results}.

In the experiment we measured \ce{Cl+} images at different values of $\beta$ for an elliptically polarized YAG pulse with the same parameters as the one used to induce 3D alignment (where $\beta = \SI{0}{\degree}$ or $\SI{90}{\degree}$). The results, displayed in \autoref{fig:beta}, are very similar to the results obtained with the linearly polarized YAG pulse and establish that 1D orientation of the $z$-axis occurs. No information about the $y$-axis can be extracted from this ion species. Images of \ce{H+} and \ce{N+} ions were also recorded. Both ion species are confined along the vertical centerline of the images which shows that 3D alignment occurs - similar to the case presented in \autoref{fig:cpc-alignment}(a3) and \autoref{fig:cpc-alignment}(b3) -- as expected. Neither of these ion species are, however, well suited to extract information about the orientation of the $y$-axis. At this point we, therefore, conclude that the CPC molecules are, at least, 3D aligned and simultaneously have their $z$-axis 1D oriented.

\section{Theoretical results and comparison with observations}
\label{sec:theory-results}

\subsection{Dc field only}
\begin{figure}[t]
 \centering
  \includegraphics[scale=0.4]{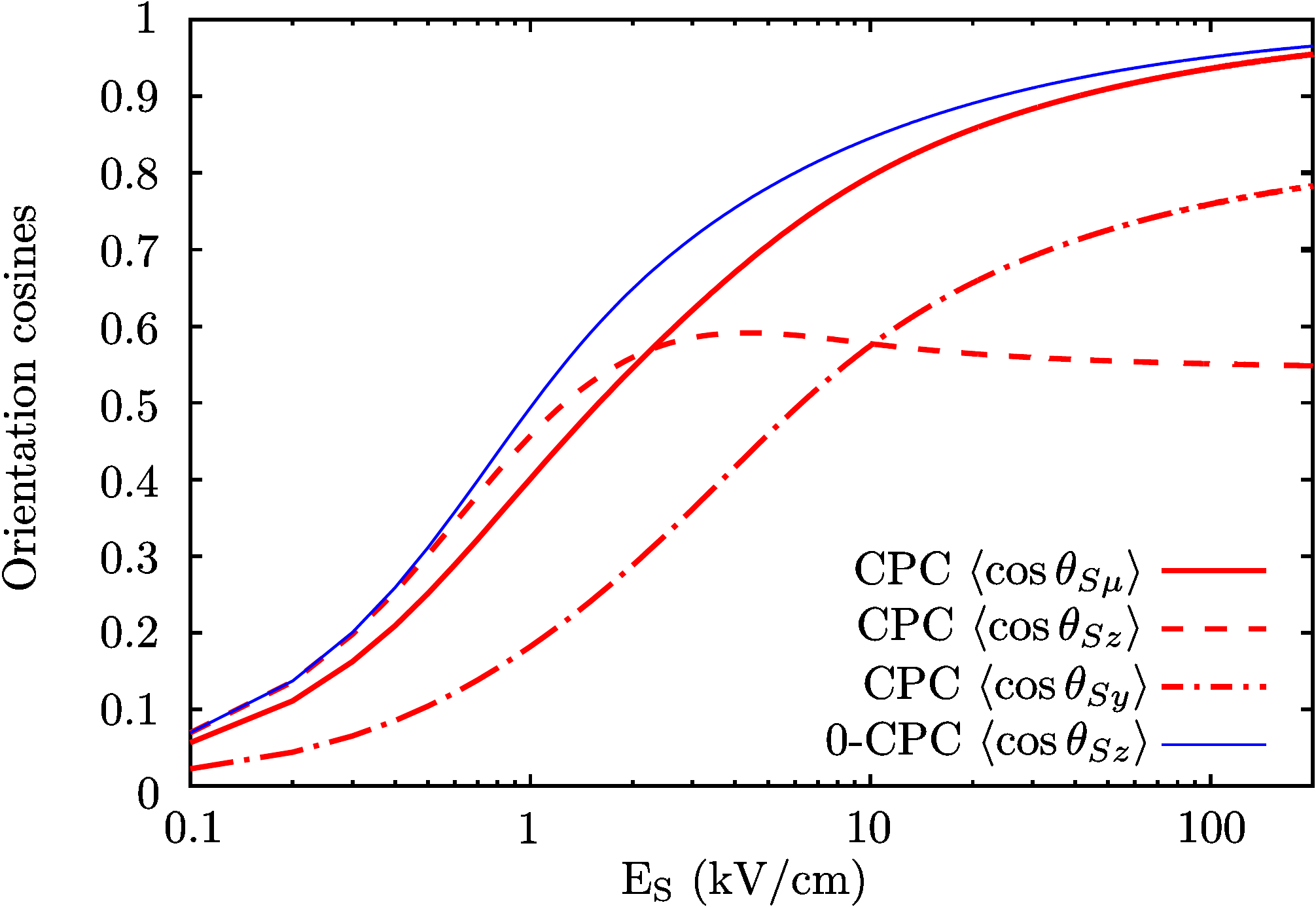}
  \caption{Expectation values
$\expected{\cos\theta_{s\mu}}$ (thick solid line), $\expected{\cos\theta_{sz}}$ (dashed line)
and $\expected{\cos\theta_{sy}}$ (dot-dashed line)
of CPC,
and $\expected{\cos\theta_{sz}}$ (thin solid line) of 0-CPC
as a function of the electric field strength $\Estatabs$.
The field configuration is    $\beta=\SI{0}{\degree}$ and $\Iyag=\SI{0}{\intensity}$.
}
  \label{fig:orientation_cosines}
\end{figure}

In this subsection we consider only the influence of a static electric field on the CPC molecule,
\ie,  $\Iyag=\SI{0}{\intensity}$.
The $\mu_z$ ($\mu_y$) term of the Stark effect interaction couples states
with different parity under inversion along the molecular $z$ ($y$) axis. As the dc field strength
is increased, the electric dipole moment $\boldsymbol{\mu}$ gets oriented along the electric field
axis. The expectation values $\expected{\cos\theta_{s\mu}}$, $\expected{\cos\theta_{sz}}$ and
$\expected{\cos\theta_{sy}}$, see \autoref{eqn:cos_smu}, \ref{eqn:cos_sz} and \ref{eqn:cos_sy},
measure the orientation of $\boldsymbol{\mu}$ and of the molecular $z$ and $y$ axes, respectively.
They are presented as a function of the dc field strength $\Estatabs$
in~\autoref{fig:orientation_cosines}. For the rotational ground state and
$\Estatabs=\SI{714}{\fieldstrength}$, we compute $\expected{\cos\theta_{s\mu}}=0.327$,
$\expected{\cos\theta_{sz}}=0.384$, and $\expected{\cos\theta_{sy}}=0.141$. In such a weak field the
orientation of the $y$-axis, corresponding to the largest dipole moment component, is smaller than
the orientation of the $z$ axis, see~\autoref{fig:orientation_cosines}, because the energy gap from
the ground state to the first level with odd parity under inversion along the $y$-axis
($\left|J_{K_aK_c	}M\right>=\left|1_{11}0\right>$) is larger than to the first level with odd parity
under the inversion along the $z$-axis ($\left|1_{01}0\right>$). When $\Estatabs$ is increased, the
hybridization of the pendular levels increases, and this trend in the orientation is inverted; we
encounter that $\expected{\cos\theta_{sz}} < \expected{\cos\theta_{sy}}$ for
$\Estatabs\gtrsim\SI{10}{k\fieldstrength}$, see~\autoref{fig:orientation_cosines}. In the
strong-dc-field regime $\lim_{\Estatabs\to\infty}\expected{\cos\theta_{s\mu}}=1$,
$\lim_{\Estatabs\to\infty}\expected{\cos\theta_{sz}}=\cos(57.1\degree)=0.543$ and
$\lim_{\Estatabs\to\infty}\expected{\cos\theta_{sy}}=\cos(32.9\degree)=0.840$.

To investigate the influence of $\mu_y$ on the dc-field orientation, we have considered a molecule
with the same rotational constants and polarizability as CPC, but with $\mu_z=2.83$~D and
$\mu_y=0$~D. When this $0$-CPC molecule is exposed to an electric field, only its $z$-axis gets
oriented along the $Z$-axis. For weak dc fields, the ground states of the CPC and 
$0$-CPC molecules show similar values of $\expected{\cos\theta_{sz}}$; we find relative differences
between $1\%$ and $5\%$ for $\SI{100}{\fieldstrength}\lesssim
\Estatabs\lesssim\SI{700}{\fieldstrength}$. By increasing $\Estatabs$, these relative differences
increase and are larger than $10\%$ for $\Estatabs\gtrsim\SI{1.2}{k\fieldstrength}$. The 0-CPC
always orients better, \ie, its orientation cosine $\expected{\cos\theta_{sz}}$ is larger than the
corresponding ones $\expected{\cos\theta_{s\mu}}$ and $\expected{\cos\theta_{sz}}$ of the CPC. Both
molecules share the same field-free energy level structure, but the $\mu_z$ and $\mu_y$ Stark
interactions couple different states, which provoke a larger orientation for 0-CPC in despite of its
smaller dipole moment. Only in the strong dc-field regime, when the pendular levels are strongly
hybridized these two systems show a similar orientation. We obtain
$\expected{\cos\theta_{s\mu}}=0.955$ for the CPC ground state and $\expected{\cos\theta_{sz}}=0.966$ for the 0-CPC ground
state at $\Estatabs=\SI{200}{k\fieldstrength}$.

\subsection{Linearly polarized laser plus dc field}
\begin{figure}[t]
 \centering
  \includegraphics[scale=0.7,angle=0]{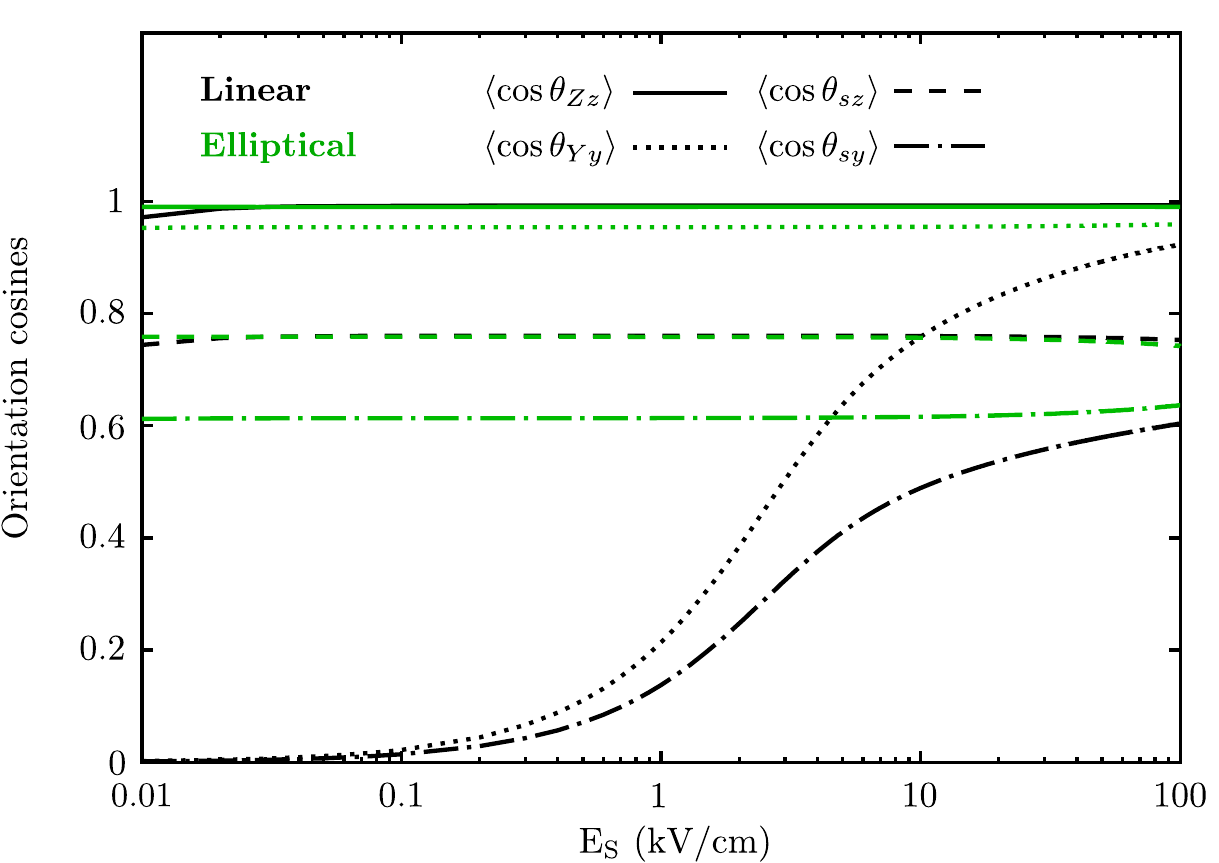}
  \caption{Expectation values for the ground state of the CPC molecule
$\expected{\cos\theta_{Zz}}$ (thick solid line),
$\expected{\cos\theta_{sz}}$ (dashed line),
$\expected{\cos\theta_{Yy}}$ (dotted line),
and $\expected{\cos\theta_{sy}}$ (dot-dashed line)
calculated
  as a function of the electric field strength $\Estatabs$
  for a linearly (black) and an elliptically (green) polarized laser pulse.
  The field configurations are $\beta=\SI{40}{\degree}$ and $\Iyag=\SI{8e11}{\intensity}$.
The ellipticity ratio is 3:1.
}
  \label{fig:orientation_cosines_linearly_polarized_laser}
\end{figure}

We now consider the molecule in a linearly polarized strong laser field, when tunneling doublets of
aligned states are formed~\cite{friedrich_enhanced_1999}. In an additional tilted weak electric
field, the terms in $\mu_z$ and $\mu_y$ in \autoref{eqn:hamil_electric} couple states in the same
doublet and
between neighbouring doublets, respectively. For the experimentally employed field-strengths, the
interaction due to the nonresonant laser field dominates. For $\Iyag = \SI{8e11}{\intensity}$, the
energy splittings of the sublevels in the lowest two pendular doublets are smaller than
$10^{-8}$~cm$^{-1}$, the energies of the two doublets differ by $0.20$~cm$^{-1}$, and the MPA is
strongly aligned along the $Z$-axis with $\expected{\cos^2\theta_{Zz}}>0.98$ for these four levels.
For a weak dc field, the $0.20$~cm$^{-1}$ energy gap between two consecutive doublets is larger than
the interaction due to the dc field: for $\Estatabs=\SI{714}{\fieldstrength}$,
$\Estatabs\mu_z=0.034$~cm$^{-1}$ and $\Estatabs\mu_y=0.053$~cm$^{-1}$. Note
that these quantities provide upper bounds to the dc field interactions because the angular
dependence in \autoref{eqn:hamil_electric} is set to $1$, which holds only for fully oriented
states. As a consequence, for weak dc fields, the coupling is only significant between states in the
same doublet and the states become oriented or antioriented along the LPFF $Z$-axis, but no
orientation of the molecular $y$-axis is achieved. This can be illustrated by a comparison between
the CPC and $0$-CPC results in this field configuration.
For $\Estatabs \ge \SI{10}{\fieldstrength}$,   these molecules present the
same mixed-field orientation of the $z$-axis $\expected{\cos\theta_{Zz}}$ with relative differences
smaller than $0.01\%$. At the experimental field regime, the mixed-field orientation of both systems
is dominated by the Stark interaction due to $\mu_z$, and the contribution of $\mu_y$ can be
neglected.
In \autoref{fig:orientation_cosines_linearly_polarized_laser}
we observe that in this regime
the  orientation cosines  along the LPFF $Z$ axis of the  CPC ground state,
\ie,  $\expected{\cos\theta_{Zz}}$ and $\expected{\cos\theta_{sz}}$,
are larger
than those along the  LPFF $Y$ axis, \ie,
$\expected{\cos\theta_{Yy}}$ and $\expected{\cos\theta_{sy}}$.
For the CPC ground state, the degrees of orientation and alignment are presented in
\autoref{tab:lineal} for $\Iyag=\SI{8e11}{\intensity}$ and $\beta=40\degree$. As $\Estatabs$ is
increased (to values $\Estatabs=\SI{5}{k\fieldstrength}$ or $\Estatabs=\SI{50}{k\fieldstrength}$ in
\autoref{tab:lineal} and up
to $\Estatabs=\SI{100}{k\fieldstrength}$
in \autoref{fig:orientation_cosines_linearly_polarized_laser}), the coupling due to $\mu_y$ is enhanced and the molecular $y$-axis gets
oriented along the LPFF $Y$-axis,
whereas  $\expected{\cos\theta_{Zz}}$ and $\expected{\cos\theta_{sz}}$
keep constant values.
In this strong electric field regime, the CPC molecule is completely 3D
oriented.
Thus, the difference between the CPC and $0$-CPC systems appears only for strong dc
fields
where they are 3D and 1D oriented, respectively. However, even in this regime, they still have the
same value of $\expected{\cos\theta_{Zz}}$.
\begin{table}
   \caption{Orientation and alignment of the ground state of the CPC molecule in a dc electric field
      and an linearly polarized YAG pulse of $\Iyag=\SI{8e11}{\intensity}$ forming an angle of
      $\beta=40\degree$.}
   \label{tab:lineal}
   \begin{ruledtabular}
      \begin{tabular}{cccc}
         $\Estatabs$ [V/cm] &$\expected{\cos^2\theta_{Zz}}$& $\expected{\cos\theta_{Zz}}$&$\expected{\cos\theta_{Yy}}$\\
         \hline
         $571$ & $0.985$& $ 0.993$&$ 0.126$\\
         $714$  &$0.985$ &$ 0.993$&$ 0.156$\\
         $5\times 10^3$ &$0.985$ &$ 0.993$&$ 0.638$\\
         $5\times 10^4$ &$0.985$ &$ 0.992$&$ 0.893$\\
      \end{tabular}
   \end{ruledtabular}
\end{table}

If the dc field is parallel to the linearly polarized laser field, the molecules will be 1D oriented
but there are no constraints in the $y$-axis. If the dc field is
perpendicular to the linearly polarized laser field, due to symmetry no orientation along this axis (the $Z$-axis)
exists. The MPA is aligned along the $Z$-axis to a degree determined by the laser intensity, \eg,
for the ground state, $\expected{\cos^2\theta_{Zz}}=0.985$ at $\Iyag =\SI{8e11}{\intensity}$. Thus,
$\mu_y$ lies close to the plane perpendicular to the $Z$-axis which includes the dc field, and the
molecular $y$-axis gets oriented along the dc field (the $Y$-axis). For $\Iyag =\SI{8e11}{\intensity}$,
$\Estatabs=\SI{714}{\fieldstrength}$, and $\beta=90\degree$, $\expected{\cos\theta_{Yy}}= 0.250$ for
the ground state. Increasing the dc field strength, this orientation is enhanced, \eg,
$\expected{\cos\theta_{Yy}}=0.731$ for $\Estatabs=\SI{5}{k\fieldstrength}$, and the molecular plane
is confined to the plane spanned by the laser field and the static field. This corresponds to 3D alignment plus 1D orientation of the $y$-axis.

\subsection{Elliptically polarized laser plus dc field}
Let us now discuss the case of an elliptically polarized laser field. The molecule becomes 3D
aligned with the most polarizable axis (the $z$-axis) confined along the $Z$-axis (the major
polarization axis) and the second most polarizable axis confined along the minor polarization axis.
Our calculation shows that $\expected{\cos^2\theta_{Zz}}> \expected{\cos^2\theta_{Yy}}$, \eg,
$\expected{\cos^2\theta_{Zz}}=0.981$ and $\expected{\cos^2\theta_{Yy}}=0.913$ for $\Iyag
=\SI{8e11}{\intensity}$. In this field configuration, the four lowest lying states with even parity
under the reflection on the LPFF $ZY$ plane belong to $4$ irreducible representations. These levels
are quasidegenerate and form a quadruplet. For $\Iyag =\SI{8e11}{\intensity}$, the energy splittings
are $7.83\times10^{-6}$~cm$^{-1}$, $7.64\times10^{-6}$~cm$^{-1}$ and $4.37\times10^{-5}$~cm$^{-1}$. In an additional electric field with
$\beta\ne0\degree,90\degree$ these states all have the same symmetry and are Stark coupled. Now,
both dc-field couplings, due to $\mu_z$ and $\mu_y$, are significantly larger than their
energy splittings. This
confinement of the molecular plane to the polarization plane is illustrated in
\autoref{fig:orientation_cosines_linearly_polarized_laser}
and  \autoref{tab:ellip} for
the CPC ground state with $\Iyag=\SI{8e11}{\intensity}$ and $\beta=40\degree$. This demonstrates that the combined action of elliptically polarized strong ac fields and weak dc-fields induces 3D orientation.

Within the adiabatic description, the orientation cosines
show a constant value
even for $\Estatabs\geq\SI{10}{\fieldstrength}$ (\autoref{fig:orientation_cosines_linearly_polarized_laser}).
Note that for the linear and elliptically polarized pulses, the orientation along the LPFF $Z$ axis is very similar, and  $\expected{\cos\theta_{Zz}}$ and $\expected{\cos\theta_{sz}}$ have very similar values in both
field configurations, see \autoref{fig:orientation_cosines_linearly_polarized_laser}.
By further increasing $\Estatabs$ above $\SI{50}{\kfieldstrength}$,
$\expected{\cos\theta_{sz}}$ and $\expected{\cos\theta_{sy}}$ present a slightly decreasing and increasing trend, respectively. They slowly approach the limits
$\lim_{\Estatabs\to\infty}\expected{\cos\theta_{sz}}=\cos(57.1\degree)=0.543$ and
$\lim_{\Estatabs\to\infty}\expected{\cos\theta_{sy}}=\cos(32.9\degree)=0.840$,
which should be reached once the Stark interaction
dominates the laser one, \ie, the mixed-field orientation in this regime resembles
 the brute-force orientation technique, but providing three-dimensional control due to the ellipticity of the ac field instead of the one-dimensional control of a pure dc field.
Let us mention that at
least two non-zero components of $\mu$ are required to achieve a 3D orientation for molecules without rotational symmetry.
\begin{table}[t]
   \caption{Orientation and alignment of the ground state of the CPC molecule in a dc electric field
      and an elliptically polarized YAG pulse with $\Iyag = \SI{8e11}{\intensity}$ and
      $\beta=40\degree$.}
   \label{tab:ellip}
   \begin{ruledtabular}
      \begin{tabular}{ccccc}
         $\Estatabs$ [V/cm]  &$\expected{\cos^2\theta_{Zz}}$& $\expected{\cos\theta_{Zz}}$&
         $\expected{\cos^2\theta_{Yy}}$& $\expected{\cos\theta_{Yy}}$\\
         \hline
         $571$ &$0.981$&$0.990$& $0.914$&$ 0.938$\\
         $714$ &$0.981$&$0.990$& $0.914$&$ 0.954$\\
         $5\times 10^3$&$0.981$ & $0.990$& $0.915$&$ 0.955$\\
         $5\times 10^4$&$0.981$ & $0.990$& $0.917$&$ 0.957$\\
      \end{tabular}
   \end{ruledtabular}
\end{table}

If the dc electric field forms an angle of $\beta=0\degree$ with the major polarization axis of the laser pulse the following cosine expectation values were obtained for $\Iyag=\SI{8e11}{\intensity}$ and $\Estatabs=\SI{10}{\fieldstrength}$: $\expected{\cos^2\theta_{Zz}}= 0.981$, $\expected{\cos^2\theta_{Yy}}=0.914$, $\expected{\cos\theta_{Sz}}=\expected{\cos\theta_{Zz}}= 0.990$, $\expected{\cos\theta_{Sy}}=8.1\times10^{-6}$. This means that the CPC molecule is 3D aligned and the $z$-axis 1D oriented -- similar to the case for higher symmetry molecules studied previously~\cite{nevo_laser-induced_2009}. The $y$-axis is equally likely to point either 'upward' or 'downward'. If $\beta$ is changed to $90\degree$ the results are: $\expected{\cos^2\theta_{Zz}}= 0.981$, $\expected{\cos^2\theta_{Yy}}=0.914$, $\expected{\cos\theta_{Sz}}= 5.4\times10^{-6}$, $\expected{\cos\theta_{Sy}}=\expected{\cos\theta_{Yy}}=0.954$, This shows that the molecules are still 3D aligned, now with the $y$-axis 1D oriented along the static electric field whereas the $z$-axis points either 'forward' or 'backward'.

Analogous features are found for the excited rotational/pendular levels. The complexity of their
field-dressed dynamics is significantly enhanced due to the large number of avoided crossings. These
avoided crossings provoke abrupt changes on their directional properties, which play an important
role for the mixed-field orientation of the molecular beam~\cite{omiste_theoretical_2011}.

For the experimentally accessed regime of dc field strengths,
$\expected{\cos\theta_{Zz}}$ is practically
independent of $\Iyag$, whereas $\expected{\cos\theta_{Yy}}$ increases until a large orientation is reached
for high intensities of elliptically polarized pulse, \eg,
 $\expected{\cos\theta_{Zz}}=0.96$ and $\expected{\cos\theta_{Yy}}=0.28$ for and
$\Iyag = \SI{5e10}{\intensity}$,  $\Estatabs=\SI{500}{\fieldstrength}$ and $\beta=40\degree$.
 Regarding the behaviour of the
orientation cosines $\expected{\cos\theta_{Zz}}$ and $\expected{\cos\theta_{Yy}}$ versus $\beta$,
three different regimes are observed: i) for weak alignment lasers, when the pendular doublets are
not yet formed, or the energy splitting between two neighbouring doublets is larger than the
dc-field interaction, $\expected{\cos\theta_{Zz}}$ or $\expected{\cos\theta_{Yy}}$ monotonically
increase with $\beta$, respectively; ii) for stronger laser fields, the energy separations between the doublets are
significantly reduced, and the orientation is independent of $\beta$; iii) if the dc-field
interaction is much larger than the laser-field interaction, the orientation in both directions reaches a
maximum at $\beta=57.1\degree$, because the effect of the static field becomes optimal at this field
configuration. In particular, this time-independent description predicts an orientation of the MPA
along the $Z$ axis independent of $\beta$ and $\Estatabs$. Thus, the smooth behaviour of
$\text{N}_{\textup{up}}/\text{N}_{\textup{total}}$ versus $\beta$ in \autoref{fig:beta} cannot be
reproduced with this theoretical treatment. Indeed, the authors have recently demonstrated that
only a time-dependent study can reproduce the intriguing physical phenomena taking place in
the mixed-field orientation experiments~\cite{nielsen_making_2012} of $C_{2v}$ symmetric molecules.

The detailed quasi-static description of the CPC molecule in mixed dc and linearly or elliptically polarized non-resonant ac fields presented here provides a solid
basis for a future time-dependent study of this system, which  is beyond the scope of this work.
Let us remark that the knowledge  of the adiabatic  energy structure in mixed-fields is required for
an  adequate interpretation of the non-adiabatic phenomena taking place in the field-dressed dynamics, such as the formation of pendular doublets~\cite{nielsen_making_2012,omiste_nonadiabatic_2012,PhysRevA.88.033416,omiste_2013}.

For completeness, we theoretically investigated the mixed-field orientation of thermal samples of
CPC, in
order to mimic the state-selection, assuming that the alignment and orientation processes are
adiabatic~\cite{omiste_theoretical_2011}. For an elliptically polarized laser with
$\Iyag=\SI{8e11}{\intensity}$, $\Estatabs=\SI{714}{\fieldstrength}$, and $\beta=40\degree$, the
molecular sample at $1$~K is strongly aligned but practically not oriented, consistent with
experimental findings~\cite{hansen_imaging_2012}: $\cost=0.949$
($\expected{\cos^2\theta_{Zz}}=0.931$,
$\expected{\cos^2\theta_{Yy}}=0.680$), $\expected{\cos\theta_{Zz}}=0.015$,
$\expected{\cos\theta_{Yy}}=0.021$
and $\text{N}_{\textup{up}}/\text{N}_{\textup{total}}=0.51$. By reducing the temperature to $0.1$~K,
the alignment is slightly improved to $\cost=0.980$ ($\expected{\cos^2\theta_{Zz}}=0.976$,
$\expected{\cos^2\theta_{Yy}}=0.868$) and the orientation is strongly increased
$\expected{\cos\theta_{Zz}}=0.36$, $\expected{\cos\theta_{Yy}}=0.43$ and
$\text{N}_{\textup{up}}/\text{N}_{\textup{total}}=0.68$. This demonstrates even that for our very cold
molecular beams ($\sim\!1$~K) the state-selection of low-energy rotational states is crucial for the
creation of orientation~\cite{holmegaard_laser-induced_2009,filsinger_quantum-state_2009}.

\section{Conclusions}
We have performed a combined experimental and theoretical investigation of mixed-field orientation
of the 6-chloropyridazine-3-carbonitrile (CPC) molecule. Our studies are motivated by the fact that this molecule represents the large class of important species where the permanent dipole moment does not coincide with any of the three principal axes of polarizability.

Experimentally we showed that the combination of an elliptically polarized laser pulse and a weak static electric field, not coinciding with either the major or the minor polarization axis of the light field, leads to 3D alignment of the molecule and 1D orientation of the most polarizable axis along the major polarization axis. The experiment is not capable of determining whether the second most polarizable axis is also oriented along the minor polarization axis but our calculation shows that it should be the case. This situation represents the most comprehensive degree of rotational control and is termed complete 3D orientation. If the elliptically polarized pulse is polarized such that the major (minor) polarization axis is parallel to the static field the molecule is 3D aligned and only the most (second most) polarizable axis is 1D oriented. Furthermore, our calculations showed that complete 3D orientation can also be achieved using a linearly
polarized laser pulse and a strong dc electric field arranged under an angle similar to the angle between the
most polarizable axis and the dipole moment.

Overall, it is clear that mixed-field orientation with appropriately polarized laser fields and weak
dc fields is an effective tool for confining how complex molecules are turned in space. Even stronger
control will be achievable in upcoming experiments combining strong dc electric fields and
linearly-polarized laser fields. The degree of angular control demonstrated  provides excellent
prospects for the recording of molecular movies of complex molecules using ion-, electron-, or
photon-imaging experiments.

\section{Acknowledgement}
We are grateful to Frank Jensen for calculating the structure, dipole moment, and polarizability
components of CPC. This work has been supported by the excellence cluster ``The Hamburg Center for
Ultrafast Imaging -- Structure, Dynamics and Control of Matter at the Atomic Scale'' of the Deutsche
Forschungsgemeinschaft, including the Mildred Dresselhaus award for R.G.F..
The work was supported by the Danish Council for Independent Research (Natural
Sciences), the Lundbeck Foundation, and the Carlsberg Foundation. Financial support by the Spanish project FIS2011-24540 (MICINN), the Grants
P11-FQM-7276 and FQM-4643 (Junta de Andaluc\'{\i}a), and the Andalusian research group FQM-207 is
gratefully appreciated. J.J.O.\ acknowledges the support of ME under the program FPU. Part of this
work was done while R.G.F.\ was visitor at the Kavli Institute for Theoretical Physics, University of
California at Santa Barbara within the program ``Fundamental Science and Applications of Ultracold
Polar Molecules'' and she gratefully acknowledges partial financial support from the National Science
Foundation grant no.\ NSF PHY11-25915.



%

\end{document}